\documentclass[conference]{IEEEtran} 
\pdfoutput=1

\IEEEoverridecommandlockouts

		\usepackage{graphicx}
		\usepackage{color}

\usepackage{cite}

\usepackage{amsmath}
\usepackage{url}
\usepackage{subfigure}

\usepackage{amssymb}
\usepackage{amsfonts}
\usepackage{lineno}

\usepackage{algorithmic}
\usepackage{algorithm}

\newtheorem{definition}{\bf Definition}

\newcommand{\sched}{\lambda}

\newcommand{\pow}{p}

\newcommand{\rate}{r}
\newcommand{\arrival}{a}
\newcommand{\channel}{h}

\newcommand{\auxSched}{\upsilon}
\newcommand{\vqSched}{\Upsilon}

\newcommand{\noise}{\sigma^2}

\newcommand*{\myfigfactor}{0.65}

\DeclareMathOperator*{\argmax}{arg\,max}
\DeclareMathOperator*{\argmin}{arg\,min}

\newcommand{\expect}{\mathbb{E}\,}

\newcommand{\vect}{\boldsymbol}

\newcommand{\set}[1]{\mathcal{#1}}
\newcommand{\size}[1]{|\set{#1}|}

\newcommand{\vectx}{\vect{x}}
\newcommand{\vecty}{\vect{y}}
\newcommand{\vectY}{\vect{Y}}

\newcommand{\one}{\mathbf{1}}
\newcommand{\zero}{\mathbf{0}}

\newcommand{\texth}[1]{\!^\text{#1}}

\newcommand{\tran}{^\dag}

\usepackage{forloop}
\newcounter{loopcntr}
\newcommand{\rpt}[2][1]{%
	\forloop{loopcntr}{0}{\value{loopcntr}<#1}{#2}%
}

\newcommand{\subgroup}[1]%
{\rlap{\smash{%
			\newcount\cnt%
			\cnt \numexpr#1\relax%
			\advance\cnt -1\relax%
			$\tabcolsep=.1em\begin{tabular}[t]{|l}\multicolumn{1}{l}{}\\%
			\rpt[\cnt]{\\}
			\\\hline\end{tabular}$%
		}}}

\begin{document}

\title{\huge Energy-Efficient Resource Management in Ultra Dense Small Cell Networks: A Mean-Field Approach
	\vspace{-8pt}
}

\author{
	\IEEEauthorblockN{Sumudu Samarakoon\IEEEauthorrefmark{1}, Mehdi Bennis\IEEEauthorrefmark{1}, Walid Saad\IEEEauthorrefmark{2}, M\'{e}rouane Debbah\IEEEauthorrefmark{3} and Matti Latva-aho\IEEEauthorrefmark{1} \\}
	\IEEEauthorblockA{\small\IEEEauthorrefmark{1}Department of Communications Engineering, University of Oulu, Finland, email: \{sumudu,bennis,matti.latva-aho\}@ee.oulu.fi \\
		\IEEEauthorrefmark{2}Wireless@VT, Bradley Department of Electrical and Computer Engineering, Virginia Tech, Blacksburg, VA, email: walids@vt.edu \\
		\IEEEauthorrefmark{3}Mathematical and Algorithmic Sciences Lab, Huawei France R\&D, Paris, France, email: merouane.debbah@huawei.fr}
	\vspace{-27pt}
	\thanks{This work is supported by the 5Gto10G project, the SHARING project under the Finland grant 128010, the U.S. National Science Foundation under Grants CNS-1460333 and CNS-1460316, and the ERC Starting Grant 305123 MORE (Advanced Mathematical Tools for Complex Network Engineering).}
}

\maketitle
\nopagebreak[4]

\begin{abstract}

	In this paper, a novel approach for joint power control and user scheduling is proposed for optimizing energy efficiency (EE), in terms of bits per unit power, in ultra dense small cell networks (UDNs).
	To address this problem, a dynamic stochastic game (DSG) is formulated between small cell base stations (SBSs).
	This game enables to capture the dynamics of both queues and channel states of the system.
	To solve this game, assuming a large homogeneous UDN deployment, the problem is cast as a \emph{mean field} game (MFG) in which the MFG equilibrium is analyzed with the aid of two low-complexity tractable partial differential equations.
	User scheduling is formulated as a stochastic optimization problem and solved using the drift plus penalty (DPP) approach in the framework of \emph{Lyapunov} optimization.
	Remarkably, it is shown that by weaving notions from Lyapunov optimization and mean field theory, the proposed solution yields an equilibrium control policy per SBS which maximizes the network utility while ensuring users' quality-of-service.
	Simulation results show that the proposed approach achieves up to $18.1\%$ gains in EE and $98.2\%$ reductions in the network's outage probability compared to a baseline model.
	
\end{abstract}

\section{Introduction}\label{sec:introduction}

The exponential growth of wireless devices and their applications during the last decade compels service providers to seek 1000x data rate by 2020 alongside improvements in capacity, reliability, energy efficiency and
latency compared to existing
systems
\cite{jnl:andrews14,online:alexiou13}.
In this regard, 5G systems are expected to be ultra-dense
in nature rendering network optimization highly complex
\cite{online:alexiou13}.
Thus,
resource management including
power control and user equipment (UE) scheduling
in ultra-dense networks (UDNs) is significantly
more challenging due to the spatio-temporal traffic demand fluctuations in the network, and the increasing overhead due to the need for coordination.
Here, uncertainties in terms of queue state information (QSI) and channel state information (CSI) as well as their evolution over time play a pivotal role in resource optimization.
Unlike sparse network deployments, optimizing UDNs based on current state-of-the-art approaches face tremendous challenges such as power control and cell deployment for interference mitigation and optimizing UE association and scheduling of large number of devices~\cite{jnl:bethanabhotla13,pap:gotsis14,pap:hui13}.
Most of these works either ignore uncertainties in QSI or focus on power control or user scheduling.

Recently, mean field games (MFGs) received significant attention in the context of cellular networks with large number of players  \cite{pap:mari12,jnl:meriaux13,book:gueant11}.
In MFGs, players make their own decisions based on their own state while abstracting other players' strategies using a mean field (MF).
As a result, the MF regime allows to cast the multi-player problem into a more tractable single player problem.
The works in \cite{pap:mari12} and \cite{jnl:meriaux13}
investigate systems
under the uncertainties of different states (QSI, CSI, battery power levels) as MFGs.
However,
these works assume either one user per cell or based on a proportional-fair (PF) scheduling baseline.

The main contribution of this paper is to propose a novel decentralized joint power control and user scheduling mechanism for ultra-dense small cell networks with large number of small cell base stations (SBSs).
Due to the severe coupling in interference, SBSs compete with each other in order to maximize their own energy efficiency (EE) in terms of transmitted bit per unit power while ensuring UEs' quality-of-service (QoS).
The problem is cast as a dynamic stochastic game (DSG) in which players are the SBSs and their actions are their control vector including transmit power and user scheduling.
Due to the non-tractability of the DSG, we study the problem in the mean field regime reflecting a very dense small cell deployment.
Thus, the solution of the DSG is obtained by solving a set of coupled partial differential equations (PDEs) known as Hamilton-Jacobi-Bellman (HJB) and Fokker-Planck-Kolmogorov (FPK)~\cite{book:gueant11}.
The UE scheduling procedure is modeled as a stochastic optimization problem
and solved
via
the Lyapunov drift plus penalty (DPP) approach~\cite{jnl:bethanabhotla13,book:neely10}.
An algorithm is
proposed in which each SBS
schedules its UEs as a function of CSI, QSI and the mean field of interferers.
Remarkably, it is shown that combining the power allocation policy obtained from MFG and the DPP-based scheduling policy enables SBSs to autonomously determine their optimal transmission parameters \emph{without coordination} with other neighboring cells.
To the best of our knowledge, \emph{this is the first work combining MFG and Lyapunov frameworks within the scope of UDNs}.

The rest of this paper is organized as follows.
Section~\ref{sec:system_model} presents the system model and formulates the DSG with finite number of players.
In Section~\ref{sec:MFG}, using the assumption of large number of players, the DSG is cast as a MFG and solved.
The solution for UE scheduling at each SBS based on the DPP framework is examined in Section~\ref{sec:formulations}.
The results are discussed in Section~\ref{sec:results}
and finally, conclusions are drawn in Section~\ref{sec:conclusion}.

\section{System Model and Problem Definition}\label{sec:system_model}

Let us  consider the downlink transmission of an ultra-dense deployment of small cells consisting of a set of SBSs $\set{B}$ using a common spectrum with bandwidth $\omega$.
SBSs serve a set of UEs
$\set{M}= \set{M}_1 \cup \ldots \cup \set{M}_{\size{B}}$ where $\set{M}_b$ is the set of UEs served by SBS $b\in\set{B}$.
For UE scheduling, we use a scheduling vector $\vect{\sched}_{b}(t)=\big[ \sched_{bm}(t) \big]_{\forall m\in\set{M}_b}$ for SBS $b$ with $\sched_{bm}(t)=1$ to denote that UE $m\in\set{M}_b$ is served by SBS $b$ at time $t$ and $\sched_{bm}(t)=0$ otherwise.
The channel gain between UE $m\in\set{M}_b$ and SBS $b$ at time $t$ is denoted by $\channel_{bm}(t)$ and an additive white Gaussian noise with zero mean and $\noise$ variance is assumed.
The instantaneous data rate of UE $m$ is given by:
\begin{equation}\label{eqn:datarate_ue}
\rate_{bm}(t) = \omega\sched_{bm}(t) \log_2 \bigg( 1 + \frac{\pow_b(t)|\channel_{bm}(t)|^2}{I_{bm}(t) + \noise} \bigg),
\end{equation}
where $\pow_b(t)\in[0,\pow_b^{\text{max}}]$ is the transmission power of SBS $b$, $|\channel_{bm}(t)|^2$ is the channel gain between SBS $b$ and UE $m$,
$I_{bm}(t)=\sum_{\forall b'\in\set{B}\setminus\{b\}}\pow_{b'}(t)|\channel_{b'm}(t)|^2$ is the interference term.

We assume that SBS $b$ sends $q_{bm}(t)$ bits to UE $m\in\set{M}_b$.
Thus, the evolution of the $b$-th SBS queue,
$\vect{q}_b(t) = \big[ q_{bm}(t) \big]_{ m\in\set{M}_b}$,
is:
\begin{equation}\label{eqn:evolution_queue}
d\vect{q}_b(t) = \vect{\arrival}_b(t) - \vect{\rate}_{b}\big(t,\vectY(t),\vect{\channel}(t)\big) dt,
\end{equation}
where $\vect{\arrival}_b(t)$
and $\vect{\rate}_{b}(\cdot)$ are the vectors of arrivals and serving data rates at SBS $b$ and
$\vect{\channel}(t) = \big[ \channel_{bm}(t) \big]_{ m\in\set{M}, b\in\set{B} }$
denotes the channel vector.
Moreover, the vector of control variables $\vectY(t)=\big[\vecty_b(t),\vecty_{-b}(t)\big]$ is defined such that $\vecty_b(t)=\big[\vect{\sched}_b(t),{\pow}_b(t)\big]$ is the SBS local control vector and $\vecty_{-b}(t)$ is the control vector of interfering SBSs.
The evolution of the channels are assumed to vary according to the following stochastic model~\cite{book:rappaport02}:
\begin{equation}\label{eqn:evolution_channels}
d\channel_{bm}(t) = G\big( t, \channel_{bm}(t)\big) dt + \zeta dz_{bm}(t),
\end{equation}
where the deterministic part $G\big( t, \channel_{bm}(t)\big)$ considers path loss and shadowing while the random part $z_{bm}(t)$ with positive constant $\zeta$ includes fast fading and channel uncertainties.
The evolution of the entire system can be described by the QSI and the CSI as per (\ref{eqn:evolution_queue}) and (\ref{eqn:evolution_channels}), respectively.
Thus, we define
$\vectx(t)= \big[ \vectx_b(t) \big]_{ b\in\set{B} } \in\set{X} $
as the state of the system at time $t$ with $\vectx_b(t)=\big[\vect{q}_b(t), \vect{\channel}_b(t) \big]$ over the state space $\set{X}=\set{X}_1 \cup \ldots \cup \set{X}_{\size{B}}$.
The feasibility set of SBS $b$'s control at state $\vectx(t)$ is defined as $\set{Y}_b(t,\vectx)=\{ {\sched}_{bm}(t)\in\{0,1\}, 0 \leq {\pow}_b(t) \leq \pow^{\texth{max}}\}$.
As the system evolves, UEs need to be scheduled at each time slot based on QSI and CSI.
The service quality of UE $m\in\set{M}_b$ is ensured such that $\bar{q}_{bm}=\lim_{t\rightarrow\infty}\frac{1}{t}\int_{0}^{t}q_{bm}(\tau) d\tau \leq \infty$.

The objective of this work is to determine the control policy per SBS $b$ which maximizes a utility function $f_b(\cdot)$ while ensuring UEs' quality of service (QoS).
Let $\bar\vectY=\lim_{t\rightarrow\infty}\frac{1}{t}\sum_{\tau=0}^{t-1}\vectY(\tau)$ be the limiting time average expectation of the control variables $\vectY(t)$.
Formally, the utility maximization problem for SBS $b$ is given as follows:
\begin{subequations}\label{eqn:spectrum_sharing_optimization}
	\begin{eqnarray}
	&\underset{\bar\vecty_b}{\text{maximize}} & f_b(\bar\vecty_b,\bar\vecty_{-b}), \\
	\label{cns:user_QoS}& \text{subject to} & \bar{q}_{bm} \leq \infty \qquad \forall m\in\set{M}_b, \\
	\label{cns:collection}& & (\ref{eqn:evolution_queue}), (\ref{eqn:evolution_channels}), \\
	\label{cns:control_all}& & \vecty_b(t)\in\set{Y}_b(t,\vectx) \qquad \forall t.
	\end{eqnarray}
\end{subequations}

Furthermore, we assume that SBSs serve their scheduled UEs for a time period of $T$.
Therefore, we use the notion of \emph{time scale separation}, hereinafter.
For SBS $b\in\set{B}$, the transmit power allocation $\pow_b(t)$ is determined for each transmission and thus, is a \emph{fast} process.
However, UE scheduling $\vect{\sched}_b(t)$ is fixed for a duration of $T$ to ensure a stable transmission.
Therefore, UE scheduling is a \emph{slower} process than power allocation.

\subsection{Dynamic stochastic game among $\size{B}$ players}\label{sec:MFG_stochastic}

We focus on finding a control policy which solves (\ref{eqn:spectrum_sharing_optimization}) over a time period $[0,T]$ for a given set of scheduled UEs considering the state transitions $\vectx(0)\rightarrow\vectx(T)$.
Therefore, we define a time-and-state-based utility for SBS $b$ as $\Gamma_b \big(0,\vectx(0)\big) = \Gamma_b \big(0,\vectx(0),\vectY(0)\big) = \expect\big[ \int_0^T f_b(\tau) d\tau \big]$.
The goal of each SBS is to maximize the above utility over $\vecty_b(\tau) = \big[\vect{\sched}^\star_b(\tau),\pow_b(\tau)\big]$ for a given UE scheduling $\vect{\sched}^\star(\tau)$ subject to the system state dynamics $d\vectx(t) = \vect{X}_t dt + \vect{X}_z d\vect{z}(t)$, $\forall b\in\set{B}, \forall m\in\set{M}$ where: 
\begin{equation*}
\vect{X}_t= \Big( \arrival_{bm}(t)-\rate_{bm}\big(t,\vectY(t),\vect{\channel}(t)\big), G\big( t, \channel_{bm}(t)\big) \Big),
\end{equation*}
and $\vect{X}_z=\text{diag}(\zero_{\size{M}},\zeta\one)$. 

As the network state evolves as a function of QSI and CSI, the strategies of SBSs need to be adaptive accordingly.
Thus, maximizing the utility $\Gamma_b \big(0,\vectx(0)\big)$ for $b\in\set{B}$ under the evolution of the network states can be modeled as a DSG.
\begin{definition}
	The formulated power control DSG for a given set of scheduled UEs is defined as
	$\set{G} = (\set{B}, \{\set{Y}_b\}_{b\in\set{B}}, \{\set{X}_b\}_{b\in\set{B}}, \{\Gamma_b\}_{b\in\set{B}})$ where:
	\begin{itemize}
		\item $\set{B}$ is the set of players which are the SBSs.
		\item $\set{Y}_b$ is the set of actions of player $b\in\set{B}$ which are the choices of transmit power $\pow_b$ for given scheduled UEs $\vect{\sched}_b$.
		\item $\set{X}_b$ is the state space of player $b\in\set{B}$ and consists of QSI $\vect{q}_b$ and CSI $\vect{h}_b$.
		\item $\Gamma_b$ is the average utility of player $b\in\set{B}$ which depends on the state transition $\vectx(t)\rightarrow\vectx(T)$ as follows:
		\begin{equation}\label{eqn:running_utility}
		\Gamma_b\big(t,\vectx(t)\big) = \Gamma_b\big(t,\vectx(t),\vectY(t)\big) = \textstyle\expect \big[ \int_t^T  f_b(\tau) d\tau \big].
		\end{equation}
	\end{itemize}
\end{definition}
The solution for the defined DSG is the Nash equilibrium (NE) defined as follows:
\begin{definition}
	The control variables $\vectY^\star(t)\in\set{Y}$ constitute a \emph{closed-loop Nash equilibrium} if,
	$$ \Gamma_b\big(t,\vectx(t),\vecty^\star_b(t),\vecty^\star_{-b}(t)\big) \geq \Gamma_b\big(t,\vectx(t),\vecty_b(t),\vecty^\star_{-b}(t)\big), $$
	$\forall\ b\in\set{B}$, $\forall\ \vectY(t)\in\set{Y}$ and $\forall\ \vectx(t)\in\set{X}$.
\end{definition}
We denote the trajectories of the utilities induced by the NE by $\vect{\Gamma}^\star\big(t,\vectx(t)\big) = \big[ \Gamma_b^\star\big(t,\vectx(t)\big) \big]_{ b\in\set{B} }$.
The existence of the NE is ensured by the existence of a joint solution $\vect{\Gamma}\big(t,\vectx(t)\big)=\big[ \Gamma_b\big(t,\vectx(t)\big) \big]_{ \forall b\in\set{B} }$ to the following $\size{B}$ coupled HJB equations~\cite{book:gueant11}:
\begin{multline*}
\textstyle \frac{\partial}{\partial {t}}[\Gamma_b\big(t,\vectx(t)\big)] + \max_{y_b(t)} \bigg[ \vect{X}_t \frac{\partial}{\partial {\vectx}}[\Gamma_b\big(t,\vectx(t)\big)] + f_b(t) \\
\textstyle  + \frac{1}{2}\text{tr}\Big( \vect{X}_z^2\frac{\partial^2}{\partial {\vectx}^2}[\Gamma_b\big(t,\vectx(t)\big)] \Big) \bigg]  = 0,
\end{multline*}
defined for each SBS $b\in\set{B}$ and tr($\cdot$) is the matrix trace operation.
Solving $\size{B}$ mutually coupled HJB equations is complex when $\size{B}>2$.
Moreover, it requires gathering QSI and CSI from all the SBSs throughout the network which incurs a tremendous amount of information exchange.
Furthermore, it is impractical for UDNs with large $\size{B}$.
In order to tackle this problem using the concept of MF, we assume $\size{B}$ is extremely large.
MF allows to approximate a stochastic differential game, by a  more tractable model.
The mean field utility of a player only depends on his own action and state, and depends on the others through a mean field.
Thus, we cast the $\size{B}$-player DSG as a MFG.

\section{Optimal Control Policy Via Mean Field}\label{sec:MFG}

As the number of SBSs becomes large ($\size{B}\rightarrow\infty$), we assume that the interference tends to be bounded in order to have non-zero rates as observed in \cite{pap:mari12,jnl:meriaux13,book:couillet11} and each SBS implements a transmission policy based on the knowledge of its own state.
SBSs in such an environment are indistinguishable from one another resulting in a continuum of players.
This allows us to simplify the solution of the $\size{B}$ HJB equations by reducing it to two equations as discussed below.

At a given time and state $\big(t,\vectx(t)\big)$ and for a scheduled UE $m$, the impact of other SBSs on the choice of a given SBS $b\in\set{B}$ appears in the interference term, where:
\begin{equation*}
I_{bm}\big(t,\vectx(t)\big) = \textstyle\sum_{\forall b'\in\set{B}\setminus\{b\}}\pow_{b'}(t)|\channel_{b'm}(t)|^2.
\end{equation*}
As the number of SBSs grows large, we assume that the interference is bounded in which a normalization factor is introduced for the channels~\cite{book:couillet11}.
Let $\eta/\size{B}$ be the normalization factor where $\eta$ is the SBS density and thus, the channel gain becomes $\channel_{bm}(t) = \frac{\sqrt{\eta} \tilde{\channel}_{bm}(t)}{\sqrt{\size{B}}}$ with $\expect[|\tilde{\channel}_{bm}(t)|^2]=1$.
Thus,
the interference can be rewritten as follows:
\begin{equation*}
I_{bm}\big(t,\vectx(t)\big) = \frac{\eta}{\size{B}} \sum_{\forall b'\in\set{B}\setminus\{b\}}\pow_{b'}(t)|\tilde{\channel}_{b'm}(t)|^2.
\end{equation*}
which ensures a bounded interference for increasing $\size{B}$ for a fixed $\eta$.

As $\size{B}\rightarrow\infty$, all SBSs yield a continuum and thus, we can focus on a generic SBS with the state $\breve{\vectx}(t)$ at time $t$.
The density of the continuum in state $\breve{\vectx}(t)$ is given by a limiting distribution $\rho\big(t,\breve{\vectx})$:
\begin{equation}\label{eqn:mass_distribution}
\rho\big(t,\breve{\vectx}) =  \lim_{\size{B}\rightarrow\infty} \frac{1}{\size{B}} \textstyle\sum\limits_{b=1}^{\size{B}} \delta\big(\vectx_b(t)=\breve{\vectx}\big),
\end{equation}
where $\delta(\cdot)$ is the \emph{Dirac delta} function.
The original problem can be reformulated as a MFG using the continuum of players and the limiting distribution of the states defined as the MF.
Therefore, the interference with respect to the MF $\vect{\rho}(t)=\big[ \rho(t,\breve{\vectx})\big]_{ x\in\set{X}}$ is given by,
\begin{equation}\label{eqn:mean_field_interference}
I\big(t,\vect{\rho}(t)\big) = {\eta} \int_{\set{X}} \pow\big(t,x\big)|\tilde{\channel}\big(t,x\big)|^2\rho(t,\breve{\vectx}) dx,
\end{equation}
with all control variables defined based on both time and state.
Note that we have omitted the subscript $b$ and $m$ since the interest is on a generic SBS.
Therefore, the utility maximization problem and the evolution of the states are reformulated for a generic SBS as follows:
\begin{subequations}\label{eqn:utility_maximization_MFG}
	\begin{eqnarray}
	&\underset{ \big(\pow(t)|\sched^\star(t)\big), \forall t\in[0,T] }{\text{maximize}} & \Gamma \big(0,\vectx(0)\big), \\
	&\text{subject to} & d\breve{\vectx}(t) = X_t dt + X_z d\vect{z}(t), \\
	& & \vecty(t)\in\set{Y}(t,\vectx) \quad \forall t\in[0,T],
	\end{eqnarray}
\end{subequations}
where 
${X}_t=\big[D\big(t,y(t)\big), G\big( t, \tilde{\channel}(t)\big) \big]$ and
 ${X}_z=\text{diag}(0,\zeta\one)$. 
Here, $D\big(t,\vecty(t)\big)=\arrival(t) - \rate(t,\vecty(t),\tilde{\channel}(t),\vect{\rho}(t)\big)$.
Similarly, the utility of a generic SBS $\Gamma\big(t,\breve{\vectx}(t)\big)$ follows (\ref{eqn:running_utility}) with the necessary modifications.
The formal definition of the MF equilibrium is as follows:
\begin{definition}
	The control vector $\vecty^\star=(\vect{\sched}^\star,\pow^\star)\in\set{Y}$ constitutes a \emph{mean-field equilibrium} if, for all $\vecty\in\set{Y}$ with the MF distribution $\vect{\rho}^\star$, it holds that,
	\begin{equation*}
	\Gamma(\vecty^\star,\vect{\rho}^\star) \geq \Gamma(\vecty,\vect{\rho}^\star).
	\end{equation*}
\end{definition}
In the MF framework, the MF equilibrium given by the solution $\big[\Gamma^\star\big(t,\breve{\vectx}(t)\big) , \rho^\star\big(t,\breve{\vectx}(t)\big)\big]$ of (\ref{eqn:utility_maximization_MFG}) is equivalent to the NE of the $\size{B}$-players DSG~\cite{book:gueant11}.
Moreover, the optimal trajectory $\Gamma^\star\big(t,\breve{\vectx}(t)\big)$ is found by applying backward induction to a single HJB equation and
the MF (limiting distribution) $\rho^\star\big(t,\breve{\vectx}(t)\big)$ is obtained by forward solving the FPK equation as follows:
\begin{equation}\label{eqn:MFG_PDEs}
\begin{cases}
\textstyle \frac{\partial}{\partial {t}}[\Gamma\big(t,\breve{\vectx}(t)\big)] + \max_{\pow(t)} \bigg[ D\big(t,\vecty(t)\big) \frac{\partial}{\partial {q}}[\Gamma\big(t,\breve{\vectx}(t)\big)] \\
\textstyle \;\quad + f(t) + \Big( G( t, \tilde{\channel}) \frac{\partial}{\partial {\tilde{\channel}}}  + \frac{\zeta^2}{2} \frac{\partial^2}{\partial {\tilde{\channel}}^2} \Big) \Gamma\big(t,\breve{\vectx}(t)\big) \bigg] = 0, \\
\textstyle \frac{\partial}{\partial {\tilde{\channel}}}\Big[G(t,\tilde{\channel})\rho\big(t,\breve{\vectx}(t)\big)\Big]  - \frac{\zeta^2}{2} \frac{\partial^2}{\partial {\tilde{\channel}}^2}[\rho\big(t,\breve{\vectx}(t)\big)] \\
\textstyle \;\quad + \frac{\partial}{\partial {q}}\Big[ D\big(t,\vecty^\star(t)\big) \rho\big(t,\breve{\vectx}(t)\big) \Big] + \partial_{t}\rho\big(t,\breve{\vectx}(t)\big)  = 0,
\end{cases}
\end{equation}
respectively.
The optimal transmit power strategy is given by,
\begin{multline}\label{eqn:optimal_strategy_MFG}
\textstyle \pow^\star(t) = \argmax_{\pow(t)} \bigg[ X_t \frac{\partial}{\partial x}[\Gamma\big(t,\breve{\vectx}(t)\big)] + f(t) \\
\textstyle  + \frac{1}{2}\text{tr}\Big( X_z^2\frac{\partial^2}{\partial x^2}[\Gamma\big(t,\breve{\vectx}(t)\big)] \Big) \bigg].
\end{multline}
Solving (\ref{eqn:MFG_PDEs})-(\ref{eqn:optimal_strategy_MFG}) yields the behavior of a generic SBS in terms of transmission power, utility and state distribution.

\section{UE Scheduling Via Lyapunov Framework}\label{sec:formulations}

\begin{figure}[t]
	\centering
	\includegraphics[width=.8\columnwidth]{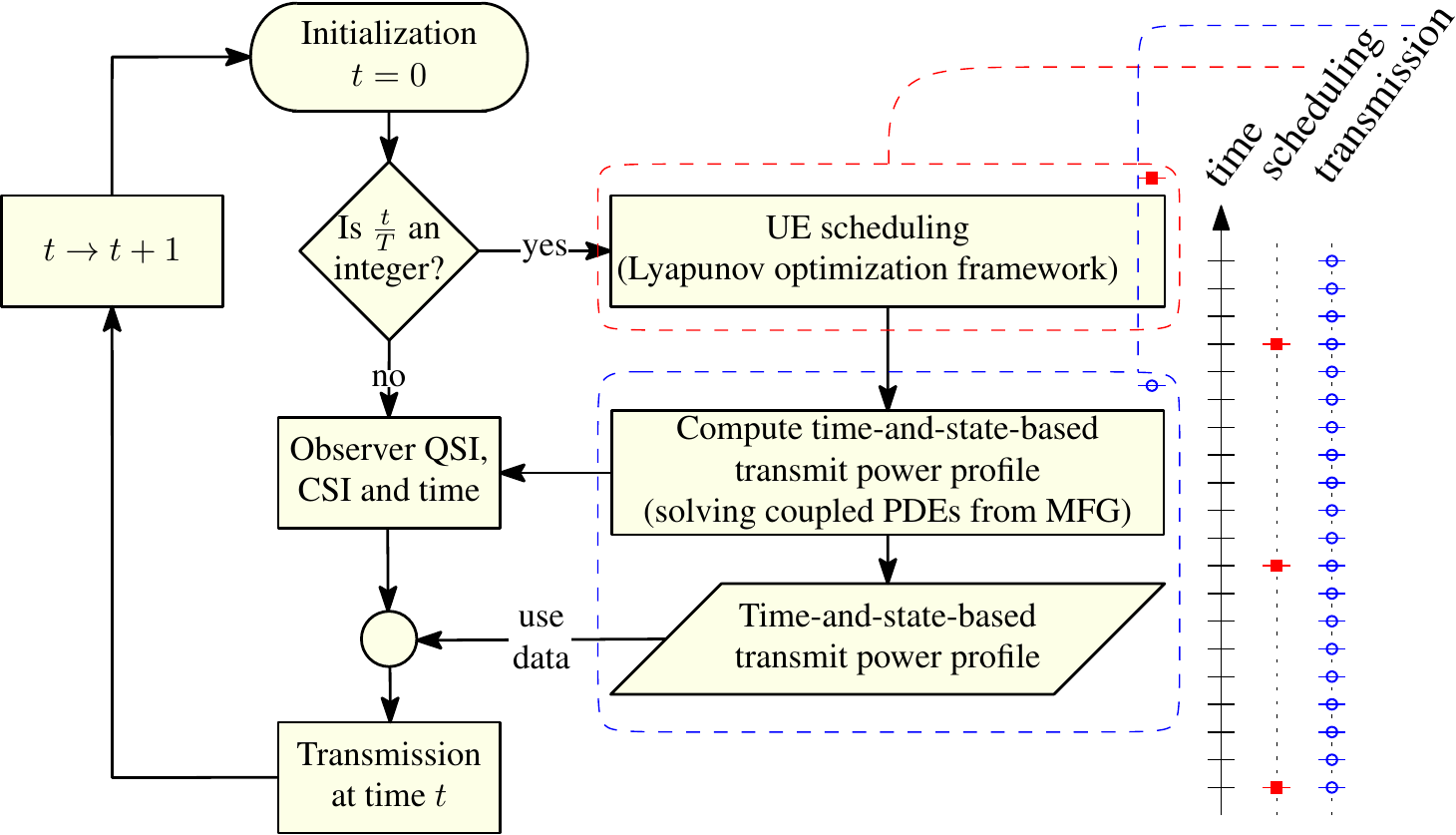}
	\caption{Inter-relation between the MFG and the Lyapunov framework.}
	\label{fig:flow_chart}
\end{figure}

By using the time scale separation, the scheduling variables are decoupled from the mean field game and thus, can be optimized separately.
Some of the baselines for UE scheduling are proportional fair (PF) scheduling in terms of rates, best-CSI based UE scheduling, and scheduling based on highest QSI.
PF scheduling ensures fairness among UEs in terms of their average rates (history) while the latter methods exploit the instantaneous CSI or QSI.
Using conventional schedulers for solving (\ref{eqn:spectrum_sharing_optimization}) fails to take advantage of the inherent CSI and QSI dynamics over space and time, thus yielding poor performance.
Therefore, we solve the original utility maximization problem of SBS $b\in\set{B}$ with respect to the scheduling variables as follows;
\begin{subequations}\label{eqn:scheduling_optimization}
	\begin{eqnarray}
	&\underset{\big( \bar{\vect{\sched}}_b|\pow^\star,\vect{\rho}^\star \big)}{\text{maximize}} & f_b(\bar\vecty_b,\bar\vecty_{-b}), \\
	\label{cns:schedule_constraints_all}&\text{subject to} & (\ref{cns:user_QoS}), (\ref{cns:collection}), \\
	\label{cns:schedules}& & \vect{\sched}_b(t) \in\set{L}(t,\vectx) \qquad \forall t.
	\end{eqnarray}
\end{subequations}
Here, the feasible set $\set{L}(t,\vectx)$ consists of all the vectors with $\sched_{bm}(t)\in[0,1]$ and $\one\tran\vect{\sched}_b(t) = 1$.
Note that the scheduling variables are relaxed from integers to real numbers for the ease of analysis.
Using the transmit power profile and the MF limiting distribution obtained from the MFG, the rate becomes:
$$\textstyle \hat{\rate}_{bm}(t)= \omega\sched_{bm}(t) \log_2 \Big( 1 + \frac{ \int_{\set{X}} \pow^\star(t-1,x)\rho^\star(t-1,x) dx }{ I\big(t-1,\vect{\rho^\star}(t-1)\big) + \noise } \Big).$$
In order to solve the stochastic optimization problem (\ref{eqn:scheduling_optimization}) per SBS $b$, the \emph{drift plus penalty} approach in Lyapunov optimization framework is applied.
The Lyapunov DPP approach decomposes the stochastic optimization problem into sub-policies that can be implemented in a distributed way.
Therefore, $\size{B}$ copies of problem (\ref{eqn:scheduling_optimization}) are locally solved at each SBS and, thus, the proposed solution applies for a large number of SBSs i.e., as $\size{B}\rightarrow\infty$.

First, a vector of auxiliary variables $\vect{\auxSched}_b(t)=\big[ \auxSched_{bm}(t) \big]_{ m\in\set{M}_b }$
is defined to satisfy the constraints (\ref{cns:schedules}).
These additional variables are chosen from a set $\set{V}$ independent from both time and state.
Thus, (\ref{eqn:scheduling_optimization}) is transformed as follows;
\begin{subequations}\label{eqn:scheduling_opt_equivalent}
	\begin{eqnarray}
	&\underset{\bar{\vect{\sched}}_b,\bar{\vect{\auxSched}}_b}{\text{maximize}} & f_b(\bar\vecty_b,\bar\vecty_{-b}), \\
	\label{cns:schedule_eq_constraints_all}&\text{subject to} &  (\ref{cns:schedule_constraints_all}), (\ref{cns:schedules}), \\
	\label{cns:aux_equivalence}& & \bar{\vect{\auxSched}}_b = \bar{\vect{\sched}}_b, \\
	\label{cns:aux_schedules}& & \vect{\auxSched}_b(t) \in\set{V} \qquad \forall t.
	\end{eqnarray}
\end{subequations}
To ensure the equality constraint (\ref{cns:aux_equivalence}), we introduce a set of virtual queues $\vqSched_{bm}(t)$ for each associated UE $m\in\set{M}_b$.
The evolution of virtual queues follows~\cite{book:neely10};
\begin{equation}
\label{eqn:virtual_queue_scheduling} \vqSched_{bm}(t+1) = \vqSched_{bm}(t) + \auxSched_{bm}(t) - \sched_{bm}(t).
\end{equation}
Consider the combined queue  $\Xi_b(t)=\big[ \vect{q}_b(t),\vect{\vqSched}_b(t) \big]$
and its quadratic Lyapunov function $L\big(\Xi_b(t)\big) = \frac{1}{2}\Xi_b\tran(t)\Xi_b(t)$.
Modifying (\ref{eqn:evolution_queue}) considering a chunk of time, the evolution of the queue of UE $m\in\set{M}_b$ can be reformulated as $q_{bm}(t+1) = \max \big(0, q_{bm}(t) + \arrival_{bm}(t) - \hat{\rate}_{bm}(t)\big)$.
Thus, one-slot drift of Lyapunov function $\Delta L = L\big(\Xi(t+1)\big)-L\big(\Xi(t)\big)$ is given by,
\begin{equation*}
\Delta L \!\! = \!\! \textstyle\frac{ \big( \vect{q}_b\tran(t+1)\vect{q}_b(t+1) - \vect{q}_b\tran(t)\vect{q}_b(t) \big)
	+ \big( \vect{\vqSched}_b\tran(t+1)\vect{\vqSched}_b(t+1) - \vect{\vqSched}_b\tran(t)\vect{\vqSched}_b(t) \big)
}{2}.
\end{equation*}
Neglecting the indexes $b,~m$ and $t$ for simplicity and using,
\begin{eqnarray*}
	&([q + \arrival - \hat{\rate}]^+)^2 &\leq q^2 + (\arrival-\hat{\rate})^2 + 2 q(\arrival-\hat{\rate}), \\
	&(\vqSched + \auxSched - \sched)^2 &\leq \vqSched^2 + (\auxSched-\sched)^2 + 2 \vqSched(\auxSched-\sched),
\end{eqnarray*}
the one-slot drift can be simplified as follows:
\begin{equation*}
\Delta L \leq K
+ \vect{q}_b\tran(t) \big( \vect{\arrival}_b(t)-\vect{\hat{\rate}}_b(t) \big)
+ \vect{\vqSched}_b\tran(t) \big( \vect{\auxSched}_b(t) - \vect{\sched}_b(t) \big),
\end{equation*}
where $K$ is a uniform bound on the term
$\big( \vect{\arrival}_b(t)-\vect{\hat{\rate}}_b(t)\big)\tran\big( \vect{\arrival}_b(t)-\vect{\hat{\rate}}_b(t)\big)
+ \big( \vect{\auxSched}_b(t)-\vect{\sched}_b(t)\big)\tran\big( \vect{\auxSched}_b(t)-\vect{\sched}_b(t)\big)$.
The conditional expected Lyapunov drift at time $t$ is defined as $\Delta\big(\Xi(t)\big) = \expect[ L\big(\Xi(t+1)\big) |  \Xi(t+1) ] -  L\big(\Xi(t)\big)$.
Let $V\leq 0$ be a parameter which controls the tradeoff between queue length and the accuracy of the optimal solution of (\ref{eqn:scheduling_opt_equivalent}) and $\vect{\sched}_b^{\texth{avg}}(t)=\frac{1}{t}\sum_{\tau=0}^{t-1}\vect{\sched}_b(\tau)$
be the current running time averages of scheduling
variables.
Introducing a penalty term
$V \nabla_{\vect{\sched}_b}\tran f\big(\vect{\sched}_b^{\texth{avg}}(t)\big) \expect [ \big(\vect{\sched}_b(t)\big) | \Xi(t)]$
to the expected drift and minimizing the upper bound of the drift DPP,
$K
+ V \nabla_{\vect{\sched}_b}\tran f\big(\vect{\sched}_b^{\texth{avg}}(t)\big) \expect [ \big(\vect{\sched}_b(t)\big) | \Xi_b(t)]
+ \expect [ \vect{q}_b\tran(t) \big( \vect{\channel}_b(t)-\vect{\hat{\rate}}_b(t) \big) | \Xi_b(t)]
+ \expect [ \vect{\vqSched}_b\tran(t) \big( \vect{\auxSched}_b(t) - \vect{\sched}_b(t) \big) | \Xi_b(t)]$,
yields the control policy of SBS $b$.
Thus, the objective of SBS $b$ is to minimize the below expression given by,
\begin{multline*}\label{eqn:join_objective}
\Big[
\overbrace{V \nabla_{\vect{\sched}_b}\tran f\big(\vect{\sched}_b^{\texth{avg}}(t)\big) \vect{\sched}_b(t)}^{\text{penalty}}
- \overbrace{\vect{q}_b\tran(t)\vect{\hat{\rate}}_b(t)}^{\text{QSI and CSI}} -
\!\! \overbrace{\vect{\vqSched}_b\tran(t)\vect{\sched}_b(t) }^{\parbox{7em}{\footnotesize Impact of virtual queue and scheduling}}
\!\!\!\!\! \Big]_{\#1} \\ + \Big[
\underbrace{\vect{\vqSched}_b\tran(t)\vect{\auxSched}_b(t)}_{\parbox{7em}{\footnotesize Impact of virtual queue and auxiliaries}}
\Big]_{\#2},
\end{multline*}
at each time $t$.
The terms $K$ and $\vect{q}_b\tran(t)\vect{\arrival}_b(t)$
are neglected since they do not depend on $\vect{\sched}_b(t)$ and $\vect{\auxSched}_b(t)$.
Note that terms $\#1$ and $\#2$ have decoupled the scheduling variables and the auxiliary variables, respectively.
Thus, the respective variables can be found independently by minimizing the individual terms.
The UE scheduling algorithm which solves (\ref{eqn:scheduling_opt_equivalent}) is given in Algorithm \ref{alg:scheduling}.

\begin{algorithm}[!t]
	\caption{UE Scheduling Algorithm Per SBS}
	\label{alg:scheduling}
	\begin{algorithmic}[1]                    
		\STATE {\bf Input:} $\vect{q}_b(t)$ and $\vect{\vqSched}_b(t)$ for $t=0$ and SBS $b\in\set{B}$.
		\WHILE{ true }
		\STATE Observation: queues $\vect{q}_b(t)$ and $\vect{\vqSched}_b(t)$, and running averages $\vect{\sched}_b^{\texth{avg}}(t)$.
		\STATE Auxiliary variables: $\vect{\auxSched}_b(t) = \argmin_{\vect{\nu}\in\set{V}} \vect{\vqSched}_b\tran(t)\vect{\nu}$.
		\STATE Scheduling: $\vect{\sched}_b(t) = \argmax_{\vect{\delta}\in\set{L}(t,\vectx)}
		\vect{q}_b\tran(t)\vect{\hat{\rate}}_b(t)
		+ \vect{\vqSched}_b\tran(t)\vect{\delta}
		- V \nabla_{\vect{\delta}}\tran f\big(\vect{\sched}_b^{\texth{avg}}(t)\big)\vect{\delta}$.
		\STATE Update: $\vect{q}_b(t+T)$, $\vect{\vqSched}_b(t+T)$ and $\vect{\sched}_b^{\texth{avg}}(t+T)$.
		\STATE $t\rightarrow t+T$
		\ENDWHILE
	\end{algorithmic}
\end{algorithm}

It is worth to mention that the resulting scheduling vector $\vect{\sched}_b(t)$ is a standard unit vector due to the affine nature of the corresponding maximization objective with respect to $\vect{\sched}_b(t)$, i.e. $\sched_{bm'}(t)=1$ only if $m'=\argmax_{m\in\set{M}} {q}_{bm}(t){\hat{\rate}}_{bm}(t)
+ {\vqSched}_{bm}(t)
- V \frac{\partial}{\partial{\sched_{bm}}}[f\big(\vect{\sched}_b^{\texth{avg}}(t)\big)]$.
Thus, scheduling a single UE at a given time instance is held, i.e. relaxing boolean schedule variable to a continuous variable does not violate the scheduling policy.
The interrelation between the MFG and the Lyapunov optimization is illustrated in Fig.~\ref{fig:flow_chart}.

\section{Numerical Results}\label{sec:results}

For the simulations, the problem needs to be simplified in order to solve the coupled PDEs using a finite element method.
We used the MATLAB PDEPE solver for the above purpose.
We assume that channels are not time-varying and thus, the state is solely defined by the QSI.
Moreover, the QSI and the scheduling time window $T$ are assumed to be normalized.
The initial
limiting distribution, $\vect{\rho}\big(0,\vectx(0)\big)$, is assumed to follow a superposition of two truncated Gaussian distributions with means $0.4$, $0.75$ and variance $0.1$, respectively.
The choice of the final utility, the boundary condition,  $\Gamma\big(T,\vectx(T)\big)=-4\exp\big(\vectx(T)\big)$ is to encourage the scheduled UE to obtain an almost empty queue by the end of its scheduled period $T$.
The arrival rate $A(t)$ for a UE is modeled as a Poisson process with mean $\bar{A}=0.2$.
The utility of a SBS at time $t$ is its EE, $r(t)/\big( p(t)+p_0 )$.
Here, $p_0$ is the fixed circuit power consumption per SBS~\cite{jnl:shuguang05}.
Due to the fact that the PDEPE solver is modeled with normalized parameters in terms of queue and time, for the purpose of simulation we assume that the transmit power is $\pow\in[0,20]$ and the variance of Gaussian noise is $\noise=1$.
Here, the SBSs and UEs are randomly distributed over the area following a uniform distribution.

\begin{figure}[!t]
	\centering
	\subfigure[Evolution of the limiting distribution for a given QSI.]{
		\includegraphics[width=\myfigfactor\columnwidth]{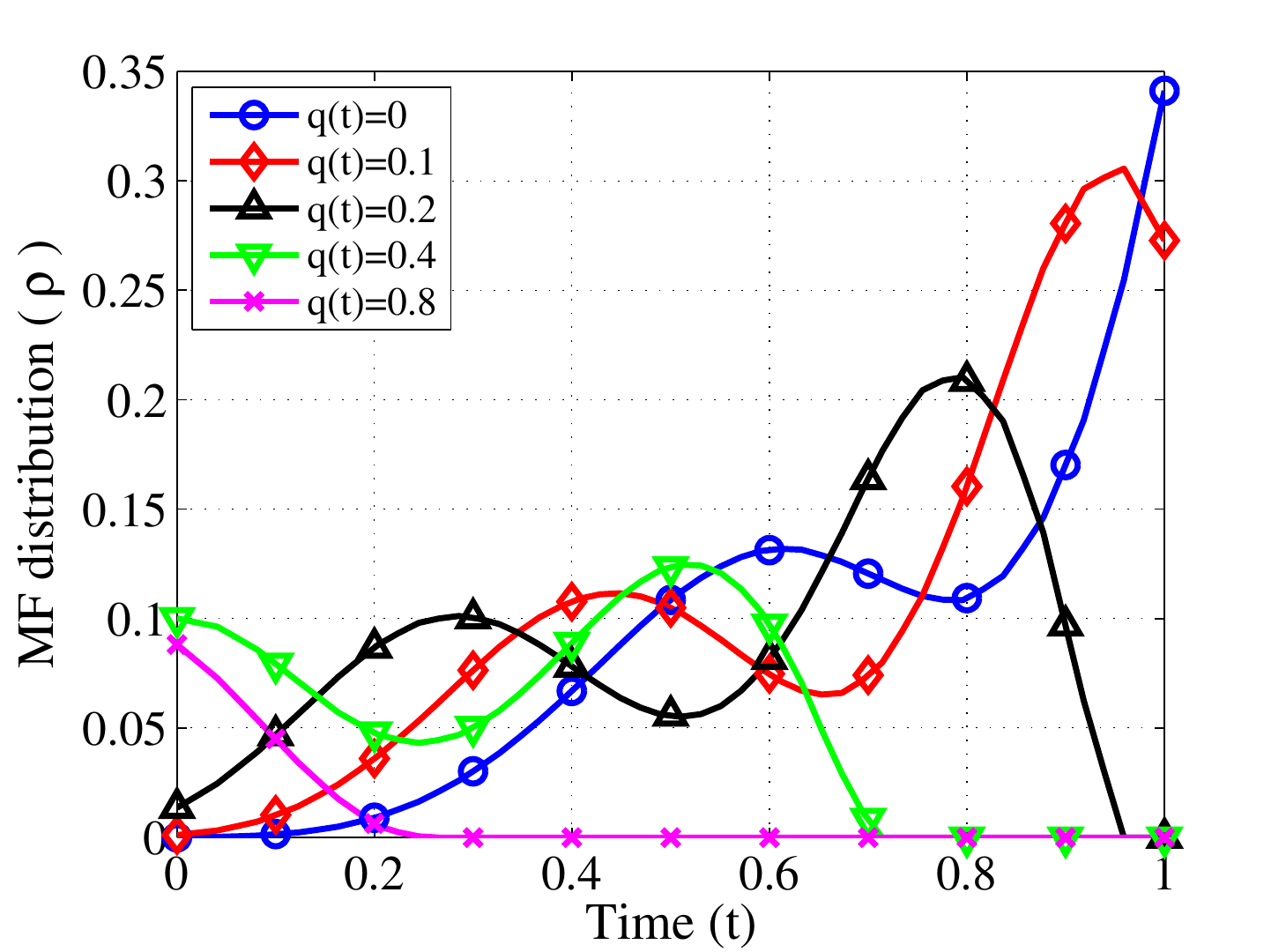}
		\label{fig:mf_distribution2D}
	}
	\hfil
	\subfigure[Transmit power at the MF equilibrium as a function of time and QSI.]{
		\includegraphics[width=\myfigfactor\columnwidth]{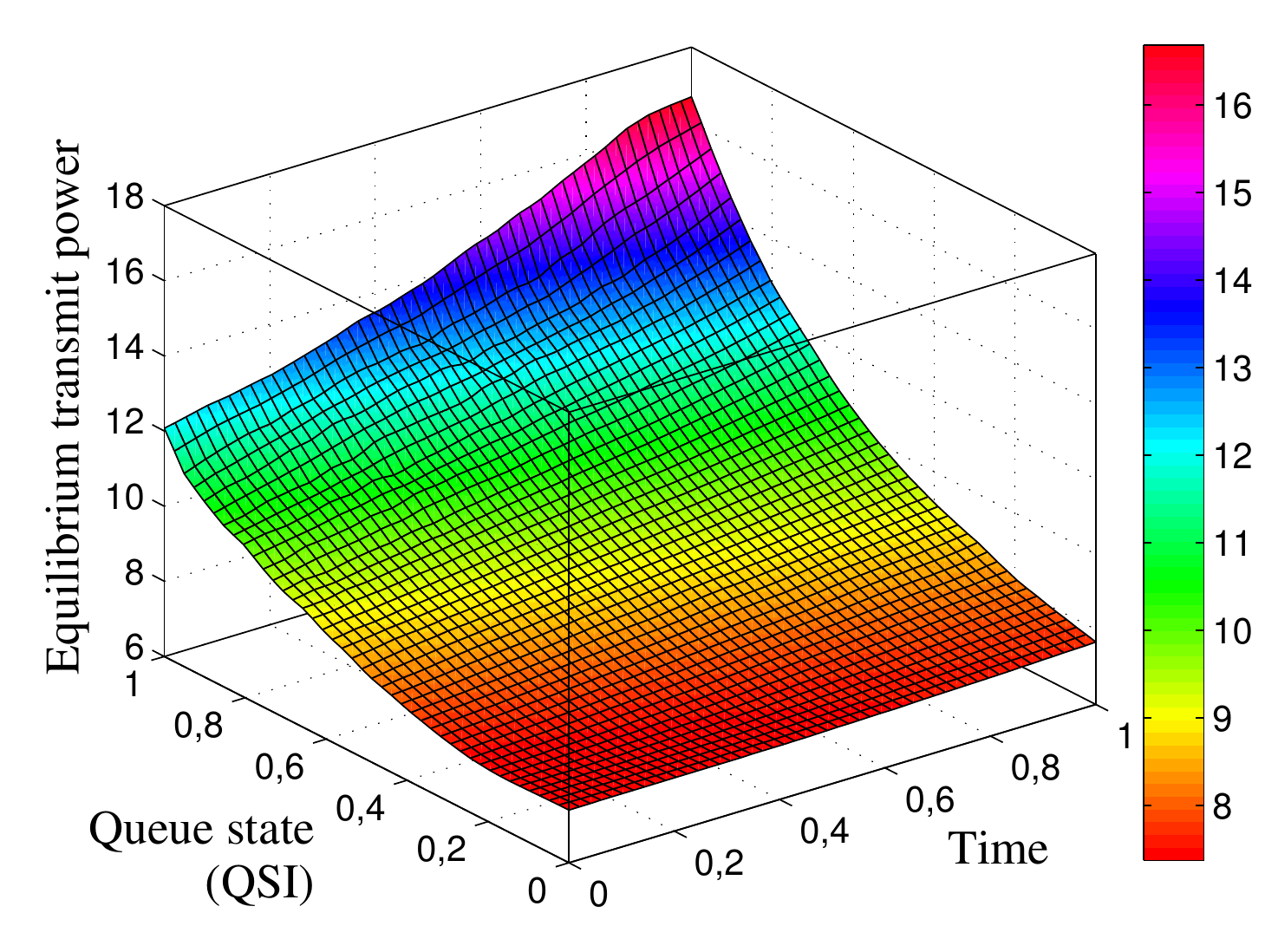}
		\label{fig:power_distribution3D}
	}
	\caption{Evolution of limiting distribution $\vect{\rho}^\star(t,q)$ and transmit power ${\pow}^\star(t,q)$ at the MF equilibrium.}
	\label{fig:mean_field}
\end{figure}

The proposed method is compared to a baseline system in which the SBS transmit powers are fixed and PF UE scheduling is used.
For a fair comparison, we consider that both models use a same average transmit power and thus, the fixed transmit power for the baseline model is set to 10 units.
The SBS density of the system is defined by the average inter-site-distance (ISD) normalized by the half of minimum ISD, i.e. minimum ISD is 2 units.
The average load per SBS is
$k=\size{M}/\size{B}$.
Once a UE is scheduled, 100 transmissions take place within the time period of $T=1$.

\subsection{Mean field equilibrium of the proposed model}

Fig.~\ref{fig:mean_field} shows the MF distribution $\vect{\rho}^\star(t,q)$, i.e. evolution of the QSI distribution of scheduled UEs over time, and transmit power policy at the MF equilibrium.
During the period of $T=1$, SBSs transmit to their scheduled UEs and
expect to achieve QSI close to zero when $t=T$ as shown in Fig.~\ref{fig:mf_distribution2D}.
It can be noted that by the end of transmission phase, the number of scheduled UEs with high QSI diminishes allowing SBSs to schedule new set of UEs for the next UE scheduling phase.
In Fig.~\ref{fig:mf_distribution2D}, we can see that the fraction of queues with $q(t)=\{0.4,0.8\}$ vanishes before the transmission duration ends.
As time evolves, the queues get empty based on the rates
prior to new arrivals
and thus, a non-monotonic increment is observed for the queue fractions with $q(t)=0$.

The transmit power policy at the MF equilibrium is shown in Fig.~\ref{fig:power_distribution3D}.
It can be observed that a higher transmit power is used when QSI is high and it is lowered at low QSI, showing the overall EE of the proposed approach.
Here, we recall that the choice of $\Gamma\big(T,\vectx(T)\big)=-4\exp\big(\vectx(T)\big)$ forces SBSs to obtain smaller QSI at $t=T$.
Thus, as time evolves, SBSs increase their transmit power for UEs with high QSI as illustrated in Fig.~\ref{fig:power_distribution3D} thereby improving the final utility.

\begin{figure}[!t]
	\centering
	\subfigure[Comparison of EE in terms of transmit bits per unit power for low and high loads $k=\{2,5\}$.]{
		\includegraphics[width=\myfigfactor\columnwidth]{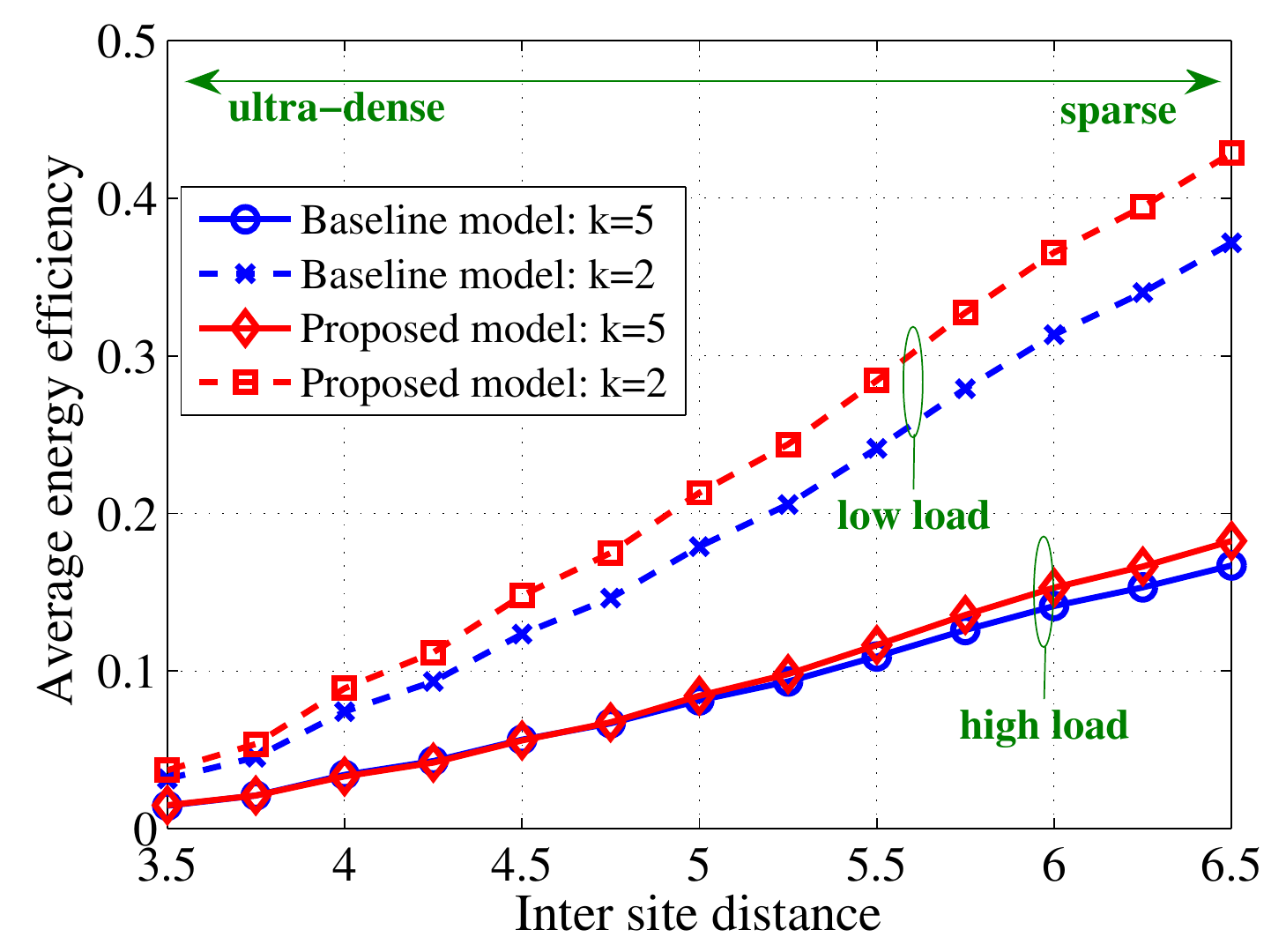}
		\label{fig:bpe_bsChange}
	}
	\hfil
	\subfigure[Comparison of outage probabilities for low and high loads $k=\{2,5\}$.]{
		\includegraphics[width=\myfigfactor\columnwidth]{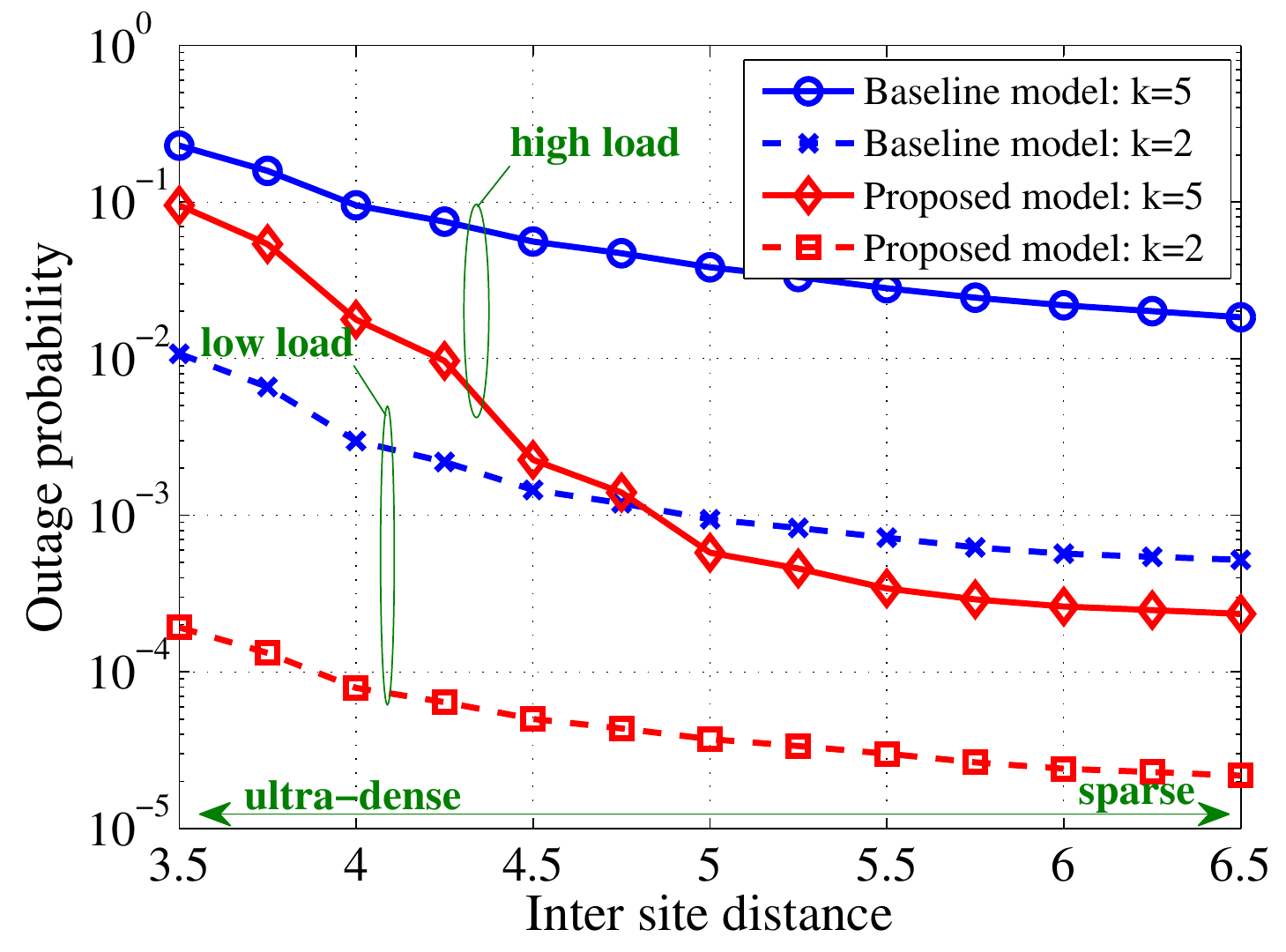}
		\label{fig:outage_bsChange}
	}
	\caption{Comparison of the behavior of EE and probability of data drops for different SBS densities.}
	\label{fig:bs_change}
\end{figure}

\subsection{Energy efficiency and outage comparisons}

In Fig.~\ref{fig:bs_change}, we show the EE of the system in terms of transmit bits per unit power and outage probability as a function of ISD.
Here, the outage probability is defined as the fraction of unsatisfied UEs whose arrivals are dropped due to limitations in queue capacity.
We can see that, for low ISD (i.e., ultra-dense scenario), due to the presence of high interference, SBSs consume high transmit power and significant amount of outages can be observed for both baseline and proposed methods.
As ISD increases, the network becomes sparse and interference reduces, resulting in increased EE and decreased outages.
The choice of fixed power for the baseline model, average power obtained by the proposed model, yields almost equal EE in both systems as illustrated in Fig.~\ref{fig:bpe_bsChange}.
However, the proposed method optimizes its power over time and QSI along DPP based UE scheduling and thus, higher energy efficiency compared to the baseline are obtained.
At low load, the average gain in EE of the proposed method is about $3.6\%$ higher compared to the baseline while it reaches up to $18.1\%$ for a high load scenario.
Although the EE gains of the proposed model are small compared to the baseline, the reductions in outage probability are significant.
From Fig.~\ref{fig:outage_bsChange}, we note that the proposed method yields $58.5\%$ and $98.2\%$ reductions in outages compared to the baseline model for both high and low loads, respectively, in UDNs.
For a sparse network, the outage reductions are $98.7\%$ and $95.8\%$ for high and low loads, respectively.

\begin{figure}[!t]
	\centering
	\subfigure[Comparison of EE for sparce and dense networks $\text{ISD}=\{5.75,3.5\}$.]{
		\includegraphics[width=\myfigfactor\columnwidth]{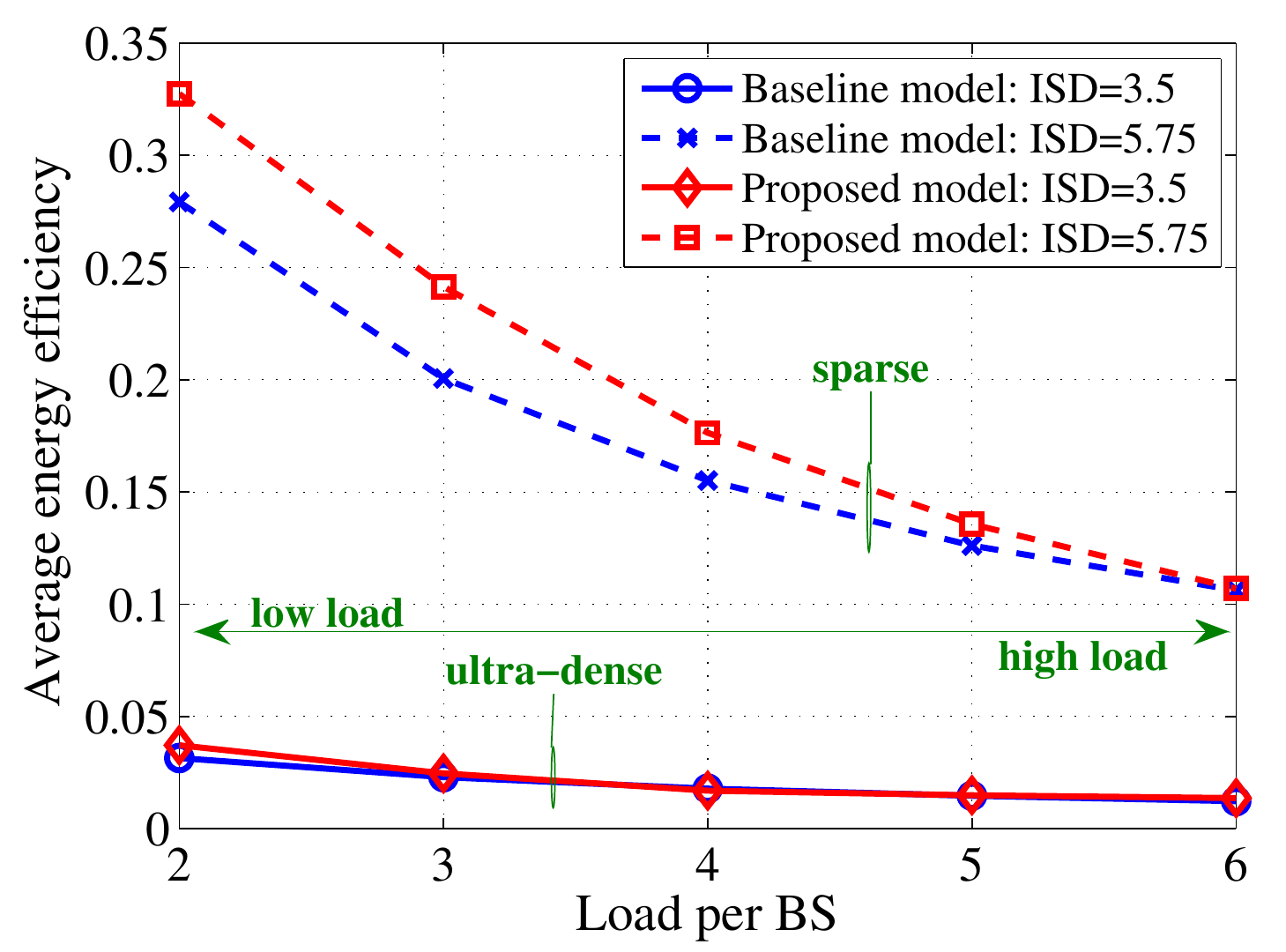}
		\label{fig:bpe_ueChange}
	}
	\hfil
	\subfigure[Comparison of average dropped data for sparce and dense networks $\text{ISD}=\{5.75,3.5\}$.]{
		\includegraphics[width=\myfigfactor\columnwidth]{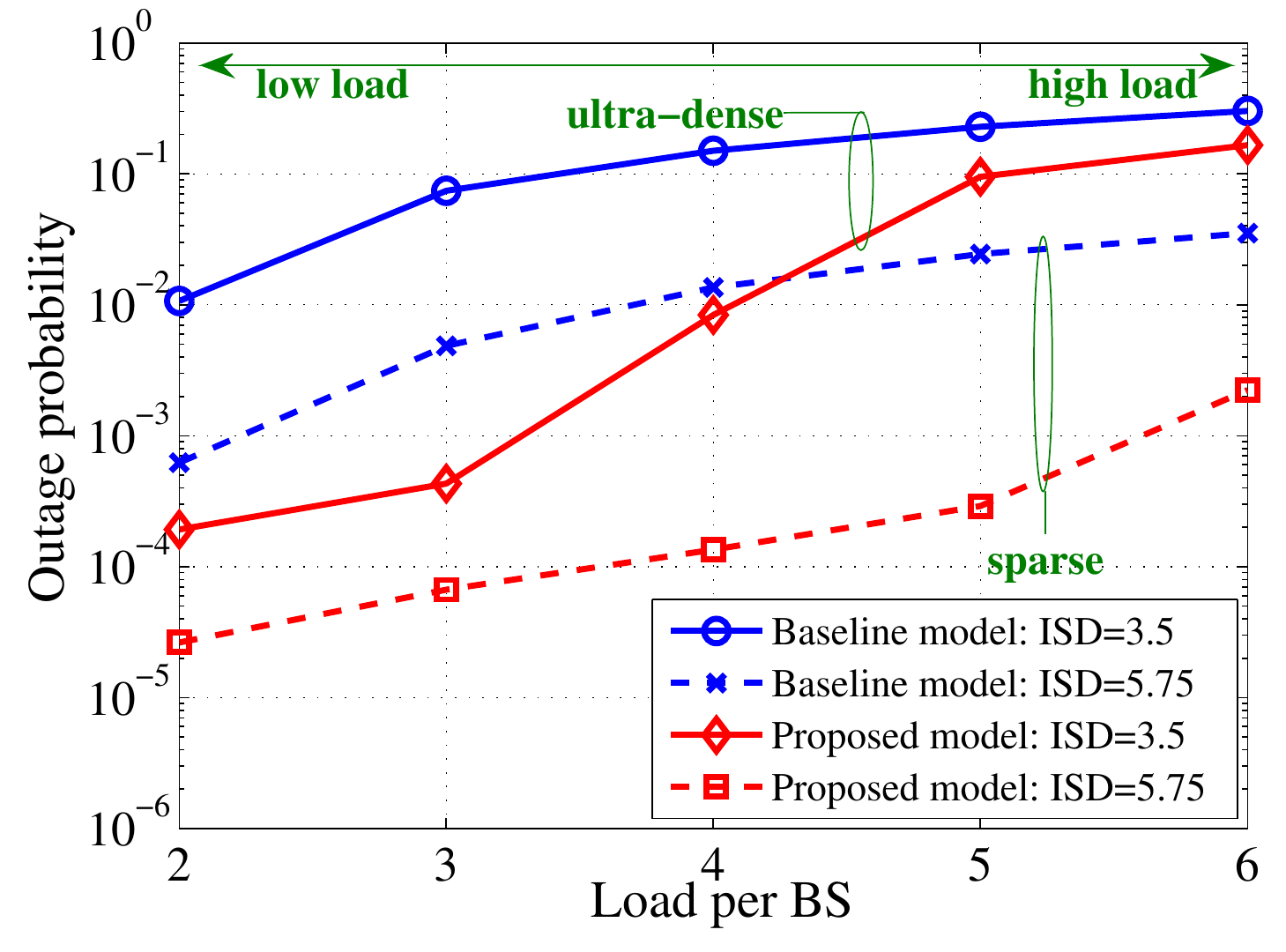}
		\label{fig:outage_ueChange}
	}
	\caption{Comparison of the behavior of EE and probability of data drops for different loads.}
	\label{fig:ue_change}
\end{figure}

Fig.~\ref{fig:ue_change} shows the EE and outrage probabilities as the loads vary.
As the load per SBS increases, UEs are scheduled with much less frequency which decreases the average rate per UE.
Therefore, a degradation in EE is observed in both methods as illustrated in Fig.~\ref{fig:bpe_ueChange}.
However, due to the adaptive nature of transmit power, the proposed method exhibits about $6.4\%$ and $11.9\%$ gains in EE compared to the baseline for ultra-dense and sparse networks, respectively.
According to Fig.~\ref{fig:outage_ueChange}, higher outage can be seen for increasing load.
These outages are low for sparse networks while significantly large for UDNs due to the increased interference and low rates.
Moreover, for a low load scenario, Fig.~\ref{fig:outage_ueChange} illustrates that the proposed method reduces the outages by $98.2\%$ and $95.7\%$ compared to the baseline model for ultra-dense and sparse networks, respectively.
As the load increases, although both models experience high outages, the proposed model displays $45.3\%$ and $93.6\%$ outage reductions compared to the baseline model with high loads and for ultra-dense and sparse scenarios, respectively.

Based on the above comparisons, it can be observed that the transmit power policy obtained by solving the MFG and the QSI aware UE scheduling using the Lyapunov framework allows SBSs to improve their EE while providing a high UEs' QoS.

\section{Conclusions}\label{sec:conclusion}
In this paper, the problem of joint power control and user scheduling for ultra-dense small cell deployment is formulated as a MFG under the uncertainties of QSI and CSI.
The goal is to maximize a time-average utility (energy efficiency in terms of bits per unit power) while ensuring users' QoS concerning outages due to queue capacity.
Under appropriate assumptions, the equilibrium of the MFG is analyzed with the aid of low-complex tractable two partial differential equations (PDEs).
While the MFG provides the optimal transmit powers, the stochastic optimization problem of user scheduling is solved via Lyapunov framework.
Numerical results have shown that the proposed method provides considerable gains in EE and massive reductions in outages compared to a baseline model.
This work is to be extended including the analytical study on the MF equilibrium along more numerical verifications.

\bibliographystyle{IEEEtran}
\bibliography{my_paper_MFG}

\end{document}